# MULTICHANNEL DATA ACQUISITION SYSTEM FOR SCINTILLATION DETECTORS OF THE EMMA EXPERIMENT

A. F. Yanin[1], V. I. Volchenko[1], L. B. Bezrukov[1], I. M. Dzaparova[1], T. Enqvist[2], H. Fynbo[4], Zh. Sh. Guliev[1], L. V. Inzhechik[1], J. Joutsenvaara[2], Y. G. Kudenko[1], P. Kuusiniemi[2], B. K. Lubsandorzhiev[1], O.V. Mineev[1], L. Olanterä[2], V. B. Petkov[1], T. Räihä[2], J. Sarkamo[2], M. Slupecki[3], W. Trzaska[3], G. V. Volchenko[1].

[1]{ Institute for Nuclear Research of RAS, Russia}
[2]{ CUPP/Pyhäsalmi, University of Oulu, Oulu Finland}
[3]{ Department of Physics, University of Jyväskylä, Finland}
[4] {Department of Physics and Astronomy, University of Århus, Denmark}
Correspondence to: A.F. Yanin  (yanin@yandex.ru)

**Annotation**

The multichannel data acquisition system is intended to be used in the EMMA experiment studying cosmic rays. The array will be in the Pihasalmi mine (central Finland) at a depth of about 85 m.

The scintillator counters (SC-1) of the array are cast plastic scintillators with a wavelength of each SC-1 detector is $12.2 \times 12.2 \times 3.0$ cm$^3$. 16 SC-1 detectors are placed in the metal case of $50.0 \times 50.0 \times 13.0$ cm$^3$ dimension. Each case, called SC-16 detector, contains electronics of preliminary processing of signals and operating mode stabilization.

The whole of the array will contain 96 detectors SC-16. It will make 1536 channels placed in three planes (48+24+24 detectors). The array will allow us to measure the time of flight of particles between SC-16 detectors and the coordinates of SC-1 fired detectors. This paper presents the function diagram of data acquisition system that includes electronic of detectors, the hodoscope pulse channels, the trigger block and VME blocks.

## 1  Introduction

The main goal of experiment EMMA consists in studying the chemical compound of cosmic rays in the knee region, with energies of $\sim 3 \times 10^{15}$ eV. The experiment will allow us to measure various characteristics of high-energy components of extended air showers (EAS) of cosmic rays. Modern microcircuits, components and ready electronic blocks (serializers, deserilizers, the FPGA, LVDS-microcircuits, chips-resistors and - capacitors, blocks of standard VME) are used in the electronics.

**2 Work of data acquisition system**

Data acquisition system (see fig. 1) will consist of 1536 small *scintillation detectors, SC-1*, of 12.2x12.2x3.0 cm$^3$ each, made of plastic [1]. Each detector SC-1 has a wavelength shifting fiber (WLS) with a multypixel avalanche photodiode (576 pixels) at one of its ends [2]. Each of the pixels can work in a mode of limited geiger breakdown when hit by photons. 16 detectors SC-1 are placed in the metal case of 50.0x50.0x13.0 cm$^3$ called *detector SC-16*. In each detector SC-16, there is a board of preliminary processing of signals, the circuit of stabilizing mode of operation for each of 16 detectors SC-1 [3], and the parallel data converter board in a fast serial channel of the LVDS-standard. Transformation is carried out by serializers. Each detector SC-16 has one digital output D$_\Sigma$ (the circuit OR on any of 16), one serial and one test (TST) channel for tuning and further expanding of possibilities. Thus, each of 96 detectors SC-16 has 3 signal connectors. Physically all detectors are distributed between the three planes as 48+24+24.

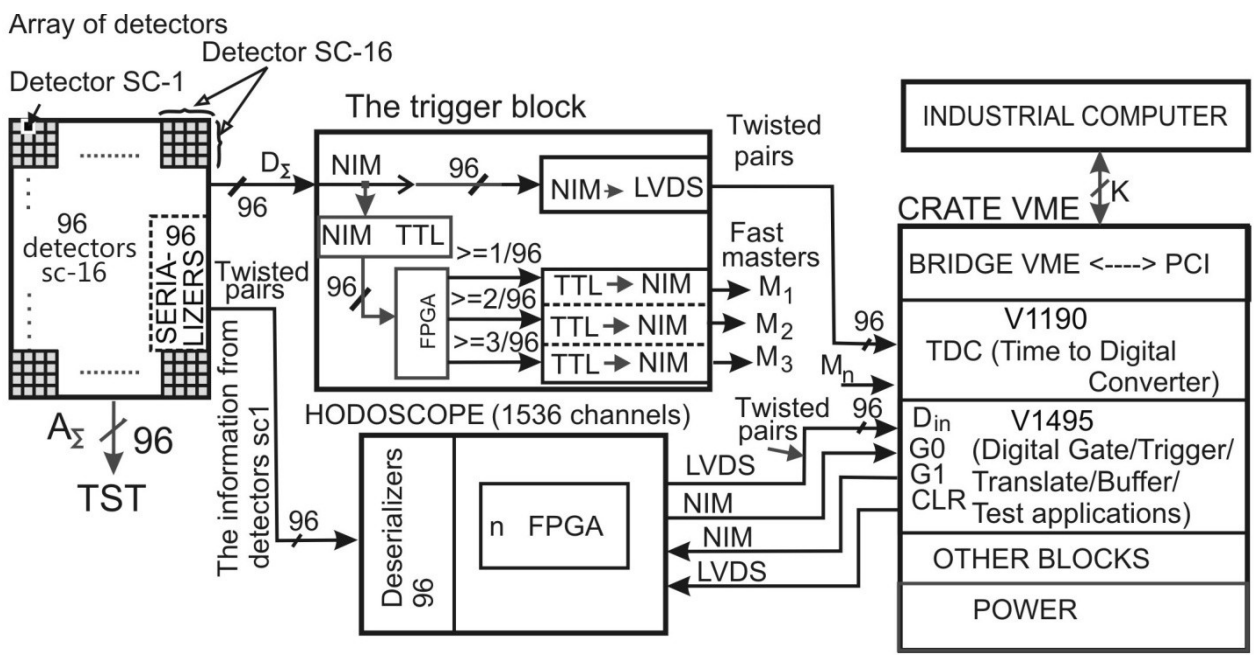

Fig. 1. Function diagram of the data acquisition system.

Digital signals from detectors are transmitted to the trigger block where the necessary master signals (M1÷M3) and the LVDS-signals are formed with help of the programmed logic integrated circuit (FPGA) and the NIM-LVDS converter respectively. The LVDS-signals are further transmitted to the industrial block TDC V1190 for measurement of relative actuation times of detectors SC-16 with a maximum resolution capability of 0,1 ns.

Serial signals of serializers are transmitted to the hodoscope using a twisted-pair cable of 10 m length. At the input of the hodoscope the deserializers restore the parallel information [4]. The data packages are transmitted with frequency of 560 MHz (maximum frequencies, up to 980 MHz, are set by external quartz). Microcircuits allow 24 bit-operation, but only 16 bits are used.

The hodoscope with help of internal or external master signal, selected by the switch, stores the information about all fired detectors SC-1 in the 16 48-bit embedded shift registers, disables input of the new information to consistently transmit the accumulated one into the computer during 47 operating clock periods G1 (the first 16-bit word is transmitted without the prior signal G1). From the hodoscope's output the data in standard LVDS are transmitted to the industrial block V1495, where they are read out by the industrial computer. Control signals G0 (command to receive the information) and G1 (to shift the information of the hodoscope) are in standard NIM. There is also a clearing LVDS-signal (CLR). In the fig. 3 the diagram of the hodoscope is presented in more detail.

Elements [1] and [2], presented in fig. 3, are decoders of number of the fired detectors of the 1$^{st}$ level. Together with the decoder of the 2$^{nd}$ level (FPGA 17 and 18) they are the decoder for the entire plane and allow us to define one (two, three) and more of the fired detectors from 768.

The fragment of functional diagram "serializer-deserializer" is presented in fig. 2.

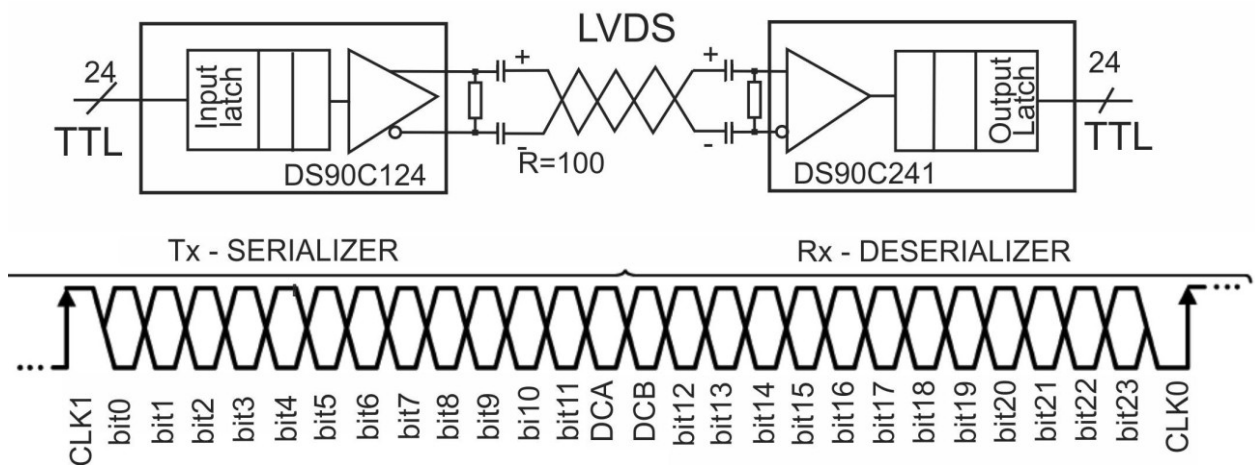

Fig. 2. The functional diagram of data transfer between serializer and deserilizer

The chips DS90C124 and DS90C241 have the 24-bit bus, but we use only 16 bits. Between the serializer and deserializer there is twisted-pair cable of 10 m length (for the plane 1). The twisted-pair cable on both ends has terminating resistors on 100 Ohm.

Configuration of FPGA is carried out by means of the configuring chip AT17LV001A (see fig. 4). This chip has a memory capacity of 1 Mbit, which is sufficient for storage of schematic diagrams of a large volume (up to 100,000 simple logic elements).

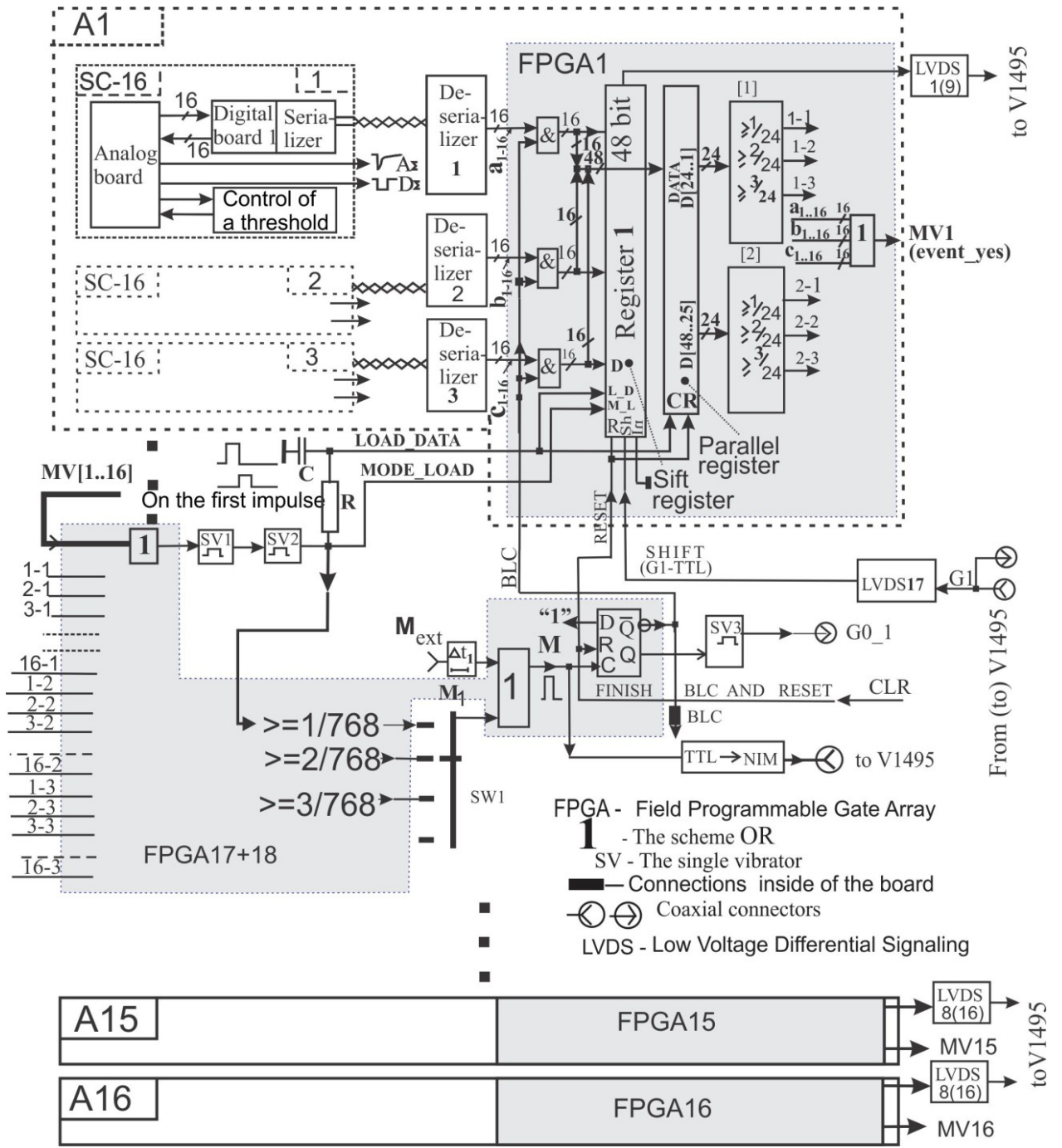

Fig. 3. Function diagram of the hodoscope jointly with the detectors.

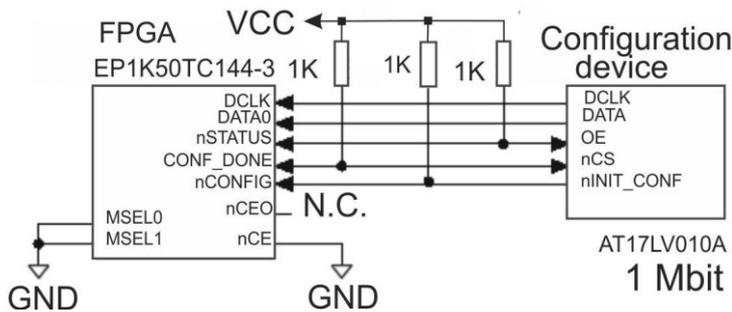

Fig. 4. The configuration diagram

## 3  Conclusions

Serializers and deserializers together with microcircuits of drivers and equalizers allow us to physically separate the detectors and hodoscope for no less than 100 m, considerably reduce quantity of cables, and to simplify installation works.

Modern electronic components used in the array's design have allowed us to lower the mass-gabarit dimensions, to reduce time of mounting the array, to flexibly reprogram if necessary the electric circuit, and to improve the operation rate and reliability of blocks.

The work is in process (final stage) and there are some additional parameters of the data acquisition system that are yet to be taken.